\documentclass[useAMS,usenatbib,usegraphicx]{mn2e}

\usepackage{amsmath}
\usepackage{url}

\def\fermi{{\it Fermi\/}}

\title[The dark knight falters]{The dark knight falters}
\author[N. Mirabal]{N. Mirabal$^{1,2}$\thanks{E-mail:
mirabal@gae.ucm.es}\\
$^{1}$Ram\'on y Cajal Fellow\\
$^{2}$ Dpto. de F\'isica At\'omica,
Molecular y Nuclear, Universidad Complutense de
Madrid, Spain\\
}

\begin{document}

\date{}

\pagerange{\pageref{firstpage}--\pageref{lastpage}} \pubyear{2012}

\maketitle

\label{firstpage}

\begin{abstract}
Tentative line emission at 111 and 129 GeV from 16 unassociated
{\it Fermi}-LAT point sources has been reported recently by
\citet{meng}. Together with similar features seen by {\it Fermi} in
a region near the Galactic Centre, the evidence has been interpreted 
as the spectral signature of dark matter annihilation or internal 
bremsstrahlung. 
Through a combination of supervised machine-learning algorithms and
archival multiwavelength observations we find
that 14 out of the 16 unassociated sources showing the line emission
in the Su \& Finkbeiner sample are most likely active galactic nuclei (AGN).
Based on this new evidence,
one must widen the range of possible solutions for the 100--140 
GeV excess to 
include a very distinct astrophysical explanation. While 
we cannot rule out a dark 
matter origin for the line emission in the Galactic Centre, we 
posit that if the detection in the Su \& Finkbeiner sample is
indeed real it might be related to  
accretion, bubble, or jet activity in nearby ($z < 0.2$) AGN.
Alternatively, given the right conditions, the similarity 
could be due to a chance occurrence caused by  
extragalactic background light (EBL) absorption. 
Or else one must concede that the features 
are an artefact of instrumental or calibration issues. 
\end{abstract}


\section{Introduction}
Frantic activity has ensued over the past few months 
following the report of an excess 
of {\it Fermi} gamma-ray events clustered
around 100 and 140 GeV in a region near the Galactic Centre \citep{weniger,
tempel,su1}, as well as 
in galaxy clusters \citep{hektor1}.
Dark matter annihilation and internal bremsstrahlung have
rapidly emerged as possible explanations \citep{bring,weniger}. 
Alternative interpretations have also been advanced 
\citep{profumo,boyarsky, aharonian}.  
More recently things have heated up even further with a tantalising 
claim of similar line emission at 111 and 129 GeV in 16  
unassociated sources detected in the Second {\it Fermi}-LAT catalogue (2FGL).
The detection could provide independent support  
for a dark matter origin for the line emission seen near the Galactic Centre
region \citep{meng}. Certainly, such coincidence might not only help us 
unlock the mysteries of
 dark matter, but it would also prove the existence of
dark matter subhaloes \citep{klypin,moore,springel}. 

In the absence of obvious flaws in the analysis, the collected evidence
has risen as a sort of
dark knight --
albeit an indirect one --
that might finally grant us non-gravitational access
to dark matter.
Intrigued by this possibility, we explore the nature of 
the 16 {\it Fermi} unassociated sources listed by \citet{meng}. 
Based on
machine-learning classification algorithms and multiwavelength examination,
we show that 14 out of the 16 unassociated {\it Fermi} sources
displaying the lines are likely gamma-ray AGN.
Therefore, rather than strengthening the argument,
the detection of an identical signal in the Su \& Finkbeiner sample
appears to disprove a dark matter origin for the {\it Fermi} 
features unless a set of very unique astrophysical conditions are met.

The paper is structured as follows. In Section \ref{sibyl} we
explain the machine-learning classifier {\it Sibyl}. 
In Section \ref{classp} we compile class prediction for the 16 unassociated 
{\it Fermi} sources listed in \citet{meng}. 
Section \ref{multi} details multiwavelength
searches for the potential counterparts of these 16 objects. Finally, we 
provide some 
interpretation in Section \ref{interp}.

\section{Sibyl}
\label{sibyl}
Confirmation of a truly unique
type of gamma-ray source 
would hint that we may have finally found the much sought-after 
dark matter subhaloes predicted by numerical simulations.  
As our base for such comparison, we use {\it Sibyl}, a Random 
Forests classifier that generates predictions on 
class memberships for unassociated {\it Fermi}-LAT sources 
using gamma-ray spectral features extracted from the
2FGL \citep{mirabal}. 
Only a brief description of {\it Sibyl} is presented here since it 
has been thoroughly covered in the literature \citep{mirabal,hassan}.
Random Forests (RF) create an ensemble of classifiers with
a tree structure, where the splits captures 
the complexity 
of the feature space among the set of training objects
\citep{breiman}.
To tag a new object, RF lets each classifier 
vote and then outputs a prediction based on the
majority of votes (P$>0.5$). RF also computes proximities
between pairs of objects and quantifies which variables are
instrumental to individual classification.
Previously we performed a similar analysis for unassociated 2FGL sources 
at $|b| \geq 10^\circ$ \citep{mirabal}. Here, we extend the coverage down to 
$|b| \geq 5^\circ$  in order to encompass the entire Su \& Finkbeiner sample. 
The classifier presented has been implemented with the
R randomForest package \citep{liaw}.

\begin{table*}
\caption{Predictions and voting percentages for the Su \& Finkbeiner sample, ordered by RA.}
\begin{tabular}{l c c l}
\hline
Source             & P$_{AGN}$ & P$_{Pulsar}$ & Prediction \\
\hline
2FGL J0341.8+3148c & 0.786 & 0.214 & AGN\\
2FGL J0526.6+2248  & 0.926 & 0.074 & AGN\\
2FGL J0555.9--4348  & 0.992 & 0.008 & AGN\\
2FGL J0600.9+3839  & 0.952 & 0.048 & AGN\\
2FGL J1240.6--7151  & 1.000 & 0.000 & AGN\\
2FGL J1324.4--5411  & 0.974 & 0.026 & AGN\\
2FGL J1335.3--4058  & 0.864 & 0.136 & AGN\\
2FGL J1601.1--4220  & 0.996 & 0.004 & AGN\\
2FGL J1639.7--5504  & 0.942 & 0.058 & AGN\\
2FGL J1716.6--0526c & 0.612 & 0.388 & - \\
2FGL J1721.5--0718c & 0.498 & 0.502 & - \\
2FGL J1730.8+5427  & 0.986 & 0.014 & AGN \\
2FGL J1844.3+1548  & 0.996 & 0.004 & AGN \\
2FGL J2004.6+7004  & 0.996 & 0.004 & AGN \\
2FGL J2115.4+1213  & 0.996 & 0.004 & AGN \\
2FGL J2351.6--7558  & 0.960 & 0.040 & AGN \\
\hline
\end{tabular}
\label{table1}
\end{table*}

As in \citet{mirabal}, we trained {\it Sibyl} using 
800 labelled AGNs (BL Lacs and flat-spectrum radio quasars only)
and 108 pulsars from the the complete \fermi\ LAT
2FGL catalogue \citep{pulsars,2lac,2lat}. 
There are additional gamma-ray classes in
the 2FGL, but since the 16 unassociated sources in \citet{meng}
lie at $|b| \geq 5^\circ$ we do not expect noticeable contamination from
Galactic plane sources. The main task thus is to find out whether 
the unassociated sample from \citet{meng} 
falls into these two categories or it is clearly 
different from these types of objects. 

During training and testing with the 908 labelled sources,
we used a total of 7 spectral features:
Index, Curve, Variability, 
and Flux Ratios ($FR_{12}$, $FR_{23}$, $FR_{34}$, and $FR_{45}$) 
\citep{mirabal}.
Assuming class bimodality, {\it Sibyl} achieves an accuracy rate of 97.1\%
based on majority voting (97.7\% for AGNs and 96.5\% for pulsars).
Inspection of the results shows that misidentifications tend to occur when 
less than 70\% of the individual classifiers 
(P$<0.7$) agree on a particular prediction. Therefore we set this 
threshold as our internal limit for a valid prediction.

\section{Application to the Su \& Finkbeiner sample}
\label{classp}
Initially, we want to examine whether the set of unassociated sources
showing line emission at 111 and 129 GeV \citep{meng}
is distinct in any way when compared to the
bulk of associated {\it Fermi} sources.
 For each of the  
16 unassociated {\it Fermi} sources listed in \citet{meng},
{\it Sibyl} provides a prediction that the object is
an AGN (P$_{AGN}$)  or a pulsar (P$_{Pulsar}$) based on 
individual votes tallied from the classifiers. We adopt a threshold
 P $>0.7$ to accept a prediction \emph{i.e.,} at least 70$\%$ of the trees
agree on the final decision. Conservatively, 
sources failing to reach the threshold remain formally unassociated and
most likely constitute interesting gamma-ray sources.
In total, {\it Sibyl} predicts
14 objects in the Su \& Finkbeiner sample to be AGN.
The resulting predictions and percentages of voting agreements
are listed in Table \ref{table1}. Only two objects 2FGL J1716.6--0526c
and 2FGL J1721.5--0718c remain without a firm
prediction. We note that both sources are 
also the only ones fitted with LogParabola functions in the Su \& Finkbeiner 
sample. But such a pair is not uncommon as there 
are at least 163 associated {\it Fermi} sources
with LogParabola best fittings in the 2FGL 
including numerous AGN, pulsars, and supernova remnants \citep{2lat}. 
Furthermore, both sources have attached caution flags to indicate
possible problems with the diffuse model that might lead
to odd spectral behaviour \citep{2lat}.

\begin{figure}                                                                 
\hfil                                                                          
\includegraphics[width=3.1in,angle=0.]{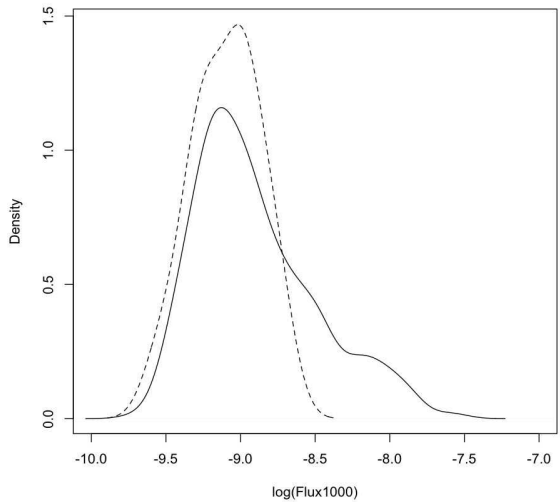}                                 
\hfil                                                                          
\caption{Kernel density plot of the 1--100 GeV photon flux   
Flux1000 (photons cm$^{-2}$ s$^{-1}$)
for the 573 unassociated sources listed in the 2FGL (solid).  
Comparison with the 16 unassociated sources in the Su \& 
Finkbeiner sample (dashed). There is no obvious difference between the
peaks of the samples.
}                                                                              
\label{figure1}                                                                
\end{figure}

Notably, we find that there are no outliers relative to 
the predicted classes among the 16 sources under scrutiny \citep{mirabal}.
Outliers correspond to cases far removed from the rest of the objects. 
To locate such cases, RF computes
the outlier measure as the inverse of the average squared proximity between
an individual object and all other objects \citep{breiman,liaw}.
Typically, outliers can be found with outlier measures greater than 10.
In the  16 sources chosen, the largest outlier measure corresponds to 1.8.
There are also no signs of
 potential dark matter subhalo candidates in 
any of the previous {\it Fermi} searches conducted
to date \citep{buckley,mirabal3,nieto,belikov,zechlin,acker12,mirabal}. 
Lastly, the photon flux distribution for the 16 unassociated sources
is compared to the overall distribution for the full
{\it Fermi}-LAT unassociated sample (573 sources) in  
Figure \ref{figure1}. Application of the Wilcoxon rank sum test
returned a p-value of $p$ = 0.07946, which indicates that the distributions
are not statistically significantly different.
Except for a slight mismatch at 
the very bright end, we find no 
obvious selection effects that could produce line emission in this particular
set.

\section{Multiwavelength examination}
\label{multi}
To the untrained  eye, it might seem overly simplistic to rely on
machine-learning algorithms to make a class prediction for a particular 
source. We simply refer the reader 
to the vast amount of research and applications of 
classification methods 
that have managed to reach 
tremendous accuracies in a variety of astrophysical subfields
\citep{bloom,richards}.
However, one must never forget that a smart computer
generated guess is no substitute for observation \footnote{Paraphrasing 
a philosophical note by Random Forests creator Leo Breiman}. 
We take this recommendation 
at heart and move the association process even further by
searching for the actual counterparts in archival multiwavelength observations 
that have partially or fully covered
the  {\it Fermi} 95\% confidence
error ellipses of the 16 unassociated sources. For this purpose, we
employ a set of
well-validated strategies \citep{mirabal2,reimer,muk}. 

The wide distribution of sources over the sky results on a  hodgepodge
of radio and X-ray observations from various existing catalogues. 
For radio counterparts, we relied on measurements from 
the Green Bank (GB6) catalogue at 4.85 GHz \citep{gregory},
the 4.85 GHZ Parkes-MIT-NRAO (PMN) survey
catalogue \citep{griffith}, the 1.4 GHz NVSS catalogue \citep{condon},
and the 843 MHz SUMSS catalogue \citep{mauch}.
In total, 
we find that 13 out of the the 16 unassociated sources have prominent 
potential radio counterparts within their {\it Fermi} 95\% confidence
error ellipses.

\begin{table*}
\caption{Potential radio and X-ray counterparts for the Su \& Finkbeiner 
sample.}
\begin{tabular}{l c c c c}
\hline
Source             & X-ray        & Counts s$^{-1}$ (0.1--2.4 keV)                        & Radio & S$_{\nu}$(mJy)\\
\hline
2FGL J0341.8+3148c &   &  & NVSS J034213+314739 & S$_{1.4 GHz}$ = 23\\
2FGL J0526.6+2248   &  &  & NVSS J052643+225337 & S$_{1.4 GHz}$ = 107\\
2FGL J0555.9--4348 &   &  & PMN J0555--4345 & S$_{4.85 GHz}$ = 61\\
2FGL J0600.9+3839 & Swift J0601.0+3838 & $(2.7 \pm 1.0) \times 10^{-3}$ & GB6 J0601+3838 & S$_{4.85 GHz}$ = 322\\
2FGL J1240.6--7125 & Swift J1240.4--7148 & $(2.6 \pm 0.1) \times 10^{-1}$ & MGPS J124021--714901 & S$_{843 MHz}$ = 18\\
2FGL J1324.4--5411 & 1RXS J132455.7--542020 & $(5.4 \pm 2.0) \times 10^{-2}$  & PMN J1325--5419  & S$_{4.85 GHz}$ = 56\\
2FGL J1335.3--4058 & & & SUMSS J133603--405758 &  S$_{843 MHz}$ = 154\\
                   & & & SUMSS J133544--410113 & S$_{843 MHz}$ = 21\\ 
                   & & & SUMSS J133535--405407 &  S$_{843 MHz}$ = 18\\
2FGL J1601.1--4220 &  & & PMN J1600--4227 & S$_{4.85 GHz}$ = 48\\
                   &  & & PMN J1600--4217 & S$_{4.85 GHz}$ =46\\
2FGL J1639.7--5504 & 1RXS J164023.6--550259 & $(2.0 \pm 0.9) \times 10^{-2}$ & &\\
2FGL J1716.6--0526c & 1RXS J171657.0--053418 & $(3.0 \pm 1.2) \times 10^{-2}$ & &\\
2FGL J1721.5--0718c & 1RXS J172147.4--071923 & $(2.3 \pm 0.9) \times 10^{-2}$ & &\\
2FGL J1730.8+5427 &           & & GB6 J1731+5429 & S$_{4.85 GHz}$ = 19\\
2FGL J1844.3+1548 & Swift J1844.4+1546 & $(7.7 \pm 0.5) \times 10^{-2}$ & GB6 J1844+1547 & S$_{4.85 GHz}$ = 76\\
2FGL J2004.6+7004 & Swift J2005.1+7004 & $(7.9 \pm 0.5) \times 10^{-2}$ & NVSS J200506+700440 & S$_{1.4 GHz}$ =7\\
2FGL J2115.4+1213 & Swift J2115.4+1218 & $(1.9 \pm 0.5) \times 10^{-2}$ & NVSS J211522+121802 & S$_{1.4 GHz}$ =16\\
2FGL J2351.6--7558 & Swift J2351.3--7600 & $(1.3 \pm 0.3) \times 10^{-2}$ & PMN J2351--7559 & S$_{4.85 GHz}$ =47\\
\hline
\end{tabular}
\label{table2}
\end{table*}

We complement the radio matches with observations from
the most ambitious X-ray program
for counterpart identification currently underway \citep{falcone}, which
aims to image
the totality of unassociated {\it Fermi} sources with the
{\it Swift} X-ray telescope. To date, nine
sources have been imaged by {\it Swift}
with times ranging from 1.1 to 19.1 ks of useful exposure.
Source extraction to identify all significant X-ray sources within
the {\it Fermi} error ellipses was performed with
{\it wavdetect}. Source positions and positional errors
were derived using {\it xrtcentroid}.  X-ray counts (0.1--2.4 keV)
were extracted from a circular
region with a 20 pixel radius (47\arcsec). The background was
extracted from an annulus with a 20 pixel (inner radius)
to 30 pixel (outer radius) around the source. Throughout, we
used {\it XSELECT} to include counts with grades 0--12.
Six out of the nine
{\it Swift} fields have single
potential counterparts within their {\it Fermi} 95\% confidence error ellipses.

\begin{figure}
\hfil
\includegraphics[width=3.1in,angle=0.]{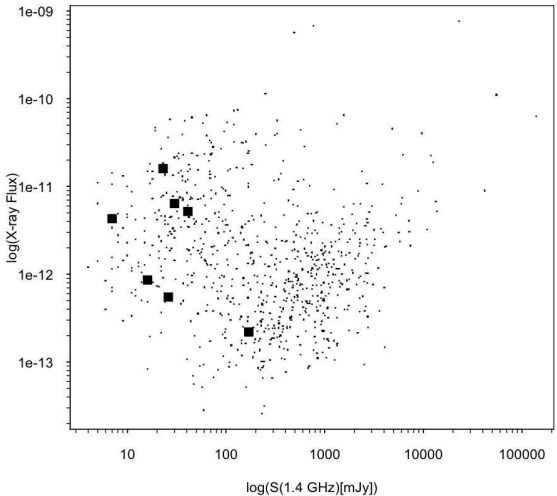}
\hfil
\caption{X-ray flux versus versus 1.4 GHz flux density
($S_{1.4 GHz}$). Small dots represent associated {\it Fermi}
AGN. The black squares mark the seven unassociated 
sources from the Su \& Finkbeiner
sample with tentative counterparts in both radio and X-rays.}
\label{figure2}
\end{figure}

The ROSAT All-Sky Survey Faint Source Catalogue \citep{voges} adds
four more potential single
X-ray counterparts to the final count.  Table \ref{table2} summarises
the counterpart candidates in each case. Of the 16 sources, seven
have both radio and X-ray tentative counterparts. Figure \ref{figure2}
shows X-ray flux versus radio flux density of associated {\it Fermi AGN}. 
Superposed are the potential seven with simultaneous radio and
X-ray counterparts.  
The results are in line with radio and X-ray counterpart flux
levels expected for typical
{\it Fermi} AGN. But we must emphasise that without dedicated spectral
classification in the optical
these must be considered solid but tentative counterparts
at this stage.

\section{Interpretation and conclusions}
\label{interp}
We have presented class predictions of the Random Forest classifier
{\it Sibyl} for 16 unassociated {\it Fermi} sources showing line emission
at 111 and 129 GeV. We find that 14 
out of 16 unassociated sources in the Su \& Finkbeiner sample 
are AGN candidates with prediction 
accuracy rates greater than 97.1\%. In addition,
we have detected 10 X-ray and 13 radio potential counterparts   
distributed over the 16 unassociated \fermi\ 95\% confidence 
error ellipses that
would be consistent with the AGN predictions.
We emphasise the word potential here as a 
more exhaustive detective work must be completed
 to confirm the appropriate counterpart
for each unassociated source.  

It was postulated that the gamma-ray lines 
among the unassociated were perhaps connected to dark matter
subhaloes dragged into the Galactic disc \citep{meng}.
However, assuming an isotropic distribution, at least
160 {\it Fermi AGN} are expected 
at $|b| \leq 10^\circ$. To date only about 100 are accounted for 
in the 2FGL \citep{2lac}. Thus, it makes astrophysical 
sense that AGN are making up an important fraction of the
Su \& Finkbeiner sample even at relatively low Galactic latitudes. 

In light of these results, the dark matter origin for the narrow gamma-ray
features observed by {\it Fermi} is in question. Were these dark
matter subhaloes \citep{baltz,diemand,kuhlen},
coincidence
between the Galactic Centre and the Su \& Finkbeiner sample would
certainly confirm a dark matter particle origin \citep{hooper}. 
However, the interpretation
changes dramatically if the unassociated sources showing an identical 
line signature are AGN, as implied by both machine-learning classifiers and
the multiwavelength arguments just presented. Dark matter could be fed
into AGN jets and the Galactic Centre, but such an explanation feels 
contrived given the hadronic
and leptonic dominance in the gamma-ray photon field \citep{hinton}. 

Instead, a distinct astrophysical mechanism 
unrelated to dark matter annihilation 
and linked to nearby AGN ($z < 0.2$ to avoid redshifted lines)  
such as accretion,  bubble \citep{meng1,profumo}, or jet \citep{su2} 
phenomenology would appear to be more logical.
However, we note that although many {\it Fermi} AGN display 
photons above $\sim$
10 GeV, only a handful of soft AGN ($\Gamma > 2$)
exhibit maximum photon energies
greater than 100 GeV at $z > 0.5$ \citep{2lac}. Consequently,
\citet{meng} might be detecting
a fiendish cluster of events imprinted by EBL absorption in the same energy
band, but completely unrelated in origin to the emission observed near the
Galactic Centre region.

Oddly enough, the lines reported by \citet{meng}
appear to be only present collectively in unassociated sources and
do not appear as pronounced among associated sources,
including well-known gamma-ray AGN \citep{meng}.
Therefore, we must also admit the possibility 
that the spectral signatures detected by {\it Fermi} 
originate from confounding instrumental or calibration 
problems \citep{hooper,hektor2,hektor3,fink}. The 
{\it Fermi} calibration team will have the final word on the matter very
soon, but
independent efforts must be made to scan the public {\it Fermi} archive 
for gamma-ray lines among 
individual AGN at $z < 0.2$, as well as in 
diffuse emission outside the Galactic plane. 

We shall hear more about this energy region
by the end of the year with the recently unveiled 
H.E.S.S. II \citep{becherini,berg}, 
and even more sensitive observations will be available later on
after completion of the Cherenkov Telescope Array  \citep{cta}. 
In the future,
a dark knight might rise again. Until then, we eagerly await 
for the final chapter of this intriguing saga.

\section*{Acknowledgments}
N.M. acknowledges support from the Spanish government 
through a Ram\'on y Cajal fellowship and the 
Consolider-Ingenio 2010 Programme under grant MultiDark CSD2009-00064. 
We thank Doug Finkbeiner  
for helpful email exchanges.
We acknowledge the use of public data from the {\it Swift} data archive. 
This research has made use of data obtained from the High Energy Astrophysics 
Science Archive Research Centre (HEASARC), provided by NASA's Goddard Space 
Flight Centre. We also thank the referee for useful suggestions and comments 
on the manuscript.

\label{lastpage}
\end{document}